\documentstyle[prl,twocolumn,aps,epsf]{revtex}

\newcommand\ket[1]{\left| #1\right\rangle}
\def\v{{\bf v}}
\def\tr{\mathop{\rm tr}}
\def\Ent{\mathop{\rm Ent}}
\def\s{{\bf s}}
\def\p{{\bf p}}
\def\vsigma{{\vec\sigma}}
\newcommand\R[1]{R^{#1}}
\newcommand\r[1]{r^{#1}}

\begin{document}
\title{Nonlocal properties of two-qubit gates and mixed states\\
and optimization of quantum computations}
\author{Yuriy Makhlin}
\address{Institut f\"ur Theoretische Festk\"orperphysik,
Universit\"at Karlsruhe, D-76128 Karlsruhe, Germany\\
Landau Institute for Theoretical Physics, 
Kosygin st. 2, 117940 Moscow, Russia\\
}

\maketitle

\begin{abstract}
Entanglement of two parts of a quantum system is a nonlocal property unaffected
by local manipulations of these parts.  It is described by quantities invariant
under local unitary transformations.  Here we present, for a system of two
qubits, a set of invariants which provides a {\it complete} description of
nonlocal properties.  The set contains 18 real polynomials of the entries of the
density matrix.  We prove that one of two {\bf mixed states} can be transformed
into the other by single-bit operations if and only if these states have equal
values of all 18 invariants.  Corresponding local operations can be found
efficiently.  Without any of these 18 invariants the set is incomplete.

Similarly, nonlocal, entangling properties of two-qubit {\bf unitary gates} are
invariant under single-bit operations.  We present a {\it complete} set of 3
real polynomial invariants of unitary gates.  Our results are useful for
optimization of quantum computations since they provide an effective tool to
verify if and how a given two-qubit operation can be performed using exactly one
elementary two-qubit gate, implemented by a basic physical manipulation (and
arbitrarily many single-bit gates).
\end{abstract}

\paragraph*{Introduction.}

Nonlocality is an important ingredient in quantum information processing, e.g. 
in quantum computation and quantum communication. Nonlocal correlations in 
quantum systems reflect entanglement between its
parts. Genuine nonlocal properties should be described in a form invariant
under local unitary operations.  In this paper we discuss such locally invariant
properties of (i) unitary transformations and (ii) mixed states of a two-qubit
system.

Two unitary transformations (logic gates), $M$ and $L$, are called locally
equivalent if they differ only by local operations:  $L=U_1MU_2$, where
$U_1,U_2\in SU(2)^{\otimes2}$ are combinations of single-bit gates on two
qubits~\cite{notation}.  A property of a two-qubit operation can be considered
nonlocal only if it has the same value for locally equivalent gates.  We
present a complete set of local invariants of a two-qubit gate:  two gates are
equivalent if and only if they have equal values of all these invariants.  The
set contains three real polynomials of the entries of the gate's matrix.  This
set is minimal:  the group $SU(4)$ of two-qubit gates is 15-dimensional and
local operations eliminate $2\dim[SU(2)^{\otimes2}]=12$ degrees of freedom;
hence any set should contain at least $15-12=3$ invariants.

This result can be used to optimize quantum computations.  Quantum algorithms
are built out of elementary quantum logic gates.  Any many-qubit quantum logic
circuit can be constructed out of single-bit and two-bit
operations~\cite{2bitUniv,Lloyd,Deutsch,9authors}.  The ability to perform 1-bit
and 2-bit operations is a requirement to any physical realization of quantum
computers.  Barenco et al.~\cite{9authors} showed that the controlled-not (CNOT)
gate together with single-bit gates is sufficient for quantum computations.
Furthermore, it is easy to prove that {\it any} two-qubit gate $M$ forms a
universal set with single-bit gates, if $M$ itself is not a combination of
single-bit operations and the SWAP-gate, which interchanges the states of two
qubits.  The efficiency of such a universal set, that is the number of
operations from the set needed to build a certain quantum algorithm, depends on
$M$.

For a particular realization of quantum computers, a certain two-qubit operation
$M$ (or a set $\exp[i{\cal H}t]$ of operations generated by a certain
hamiltonian) is usually considered elementary.  It can be performed by a basic
physical manipulation (switching of one parameter or application of a pulse).
Then the question of optimization arises:  what is the most economical way to
perform a particular computation, i.e.  what is the minimal number of elementary
steps?  In many situations two-qubit gates are more costly than single-bit gates
(e.g., they can take a longer time, involve complicated manipulations or
stronger additional decoherence), and then only the number of two-bit gates
counts.

The simplest and important version of this question is how a given two-bit gate
$L$ can be performed using the minimal number of elementary two-bit gates $M$.
In particular, when is it sufficient to employ $M$ only once~\cite{LossOptim}?  
This is
the case when the two gates are locally equivalent, and computation of
invariants gives an effective tool to verify this.  Moreover, if $M$ and $L$ are
equivalent, a procedure presented below allows to find efficiently single-bit
gates $U_1$, $U_2$, which in combination with $M$ produce $L$.  If one
elementary two-bit step is not sufficient, one can ask how many are needed.
Counting of dimensions suggests that two steps always suffice.  However, this is
not true for some gates $M$ (bad entanglers).

A related problem is that of local invariants of quantum states.  A mixed state
is described by a density matrix $\hat\rho$.  Two states are called locally
equivalent if one can be transformed into the other by local operations:
$\hat\rho\to U^\dagger\hat\rho U$, where $U$ is a local gate.  Apparently, the
coefficients of the characteristic polynomials of $\hat\rho$ and of the reduced
density matrices of two qubits are locally invariant.  The method developed in
Refs.~\cite{Rains,Grassl}, in principle, allows to compute all invariants.  For
a two-qubit system counting of dimensions shows that there should be 9
functionally independent invariants of density matrices with unit trace, but
additional invariants may be needed to resolve a remaining finite number of
states.  A set of 20 invariants was presented in Ref.~\cite{Grassl}.  However it
was not clear if this set was complete, i.e.  two states with the same
invariants are always locally equivalent.  Here we present a {\it complete} set
of 18 polynomial invariants.  We prove that two states are locally equivalent if
and only if all 18 invariants have equal values in these states.  Hence, any
nonlocal characteristic of entanglement is a function of these 
invariants~\cite{poly}.  We also show that no subset of this set is complete.

To demonstrate applications of our results, we discuss in the last section which
2-bit operations can be constructed out of only one elementary 2-bit gate for
Josephson charge qubits~\cite{Our_PRL,Our_Nature} or for qubits based on spin
degrees of freedom in quantum dots~\cite{Loss}.

\paragraph*{Single-bit gates as orthogonal matrices.}
The following result is used below to classify two-qubit gates.

{\sl Theorem 1.}
Single-qubit gates with unit determinant are represented by real 
orthogonal matrices in the Bell basis $\frac{1}{\sqrt{2}}(\ket{00}+\ket{11})$,
$\frac{i}{\sqrt{2}}(\ket{01}+\ket{10})$,
$\frac{1}{\sqrt{2}}(\ket{01}-\ket{10})$,
$\frac{i}{\sqrt{2}}(\ket{00}-\ket{11})$.

The transformation of a matrix $M$ from the standard basis of states 
$\ket{00},\ket{01},\ket{10},\ket{11}$ into the Bell basis is described as $M\to 
M_B=Q^\dagger MQ$, where
$$
Q=\frac{1}{\sqrt{2}}
\left(\begin{array}{cccc}
1&0&0&i\\
0&i&1&0\\
0&i&-1&0\\
1&0&0&-i
\end{array}\right)
\,.
$$

{\sl Proof.}  We introduce a measure of entanglement in a pure 2-qubit state
$\psi_{\alpha\beta}$, a quadratic form $\Ent\psi$, which in the standard basis
is defined as $\Ent\psi = \det\hat\psi = \psi_{00}\psi_{11} -
\psi_{01}\psi_{10}$.  This quantity is locally invariant.  Indeed, a single-bit
gate $W_1\otimes W_2$ transforms $\hat\psi$ into $W_1\hat\psi W_2^T$, preserving
the determinant.

Under a unitary operation $V$ the matrix of this form transforms as $\hat{\Ent}
\to V^T \hat{\Ent}\; V$.  In the Bell basis it is proportional to the identity
matrix.  Since local gates preserve this form, they are given by orthogonal
matrices in this basis.  As unitary and orthogonal they are also real.  Thus
local operations form a subgroup of the group $SO(4,{\bf R})$ of real orthogonal
matrices.  This subgroup is 6-dimensional, and hence coincides with the group.
$\Box$

\paragraph*{Classification of two-qubit gates.}
Our result for unitary gates is expressed in terms of $M_B$.

{\sl Theorem 2.}  The complete set of local invariants of a two-qubit gate $M$,
with $\det M=1$, is given by the set of eigenvalues of the matrix $M_B^TM_B$.

In other words, two 2-bit gates with unit determinants, $M$ and $L$, are 
equivalent up to single-bit operations iff the spectra of $M_B^TM_B$ and 
$L_B^TL_B$ coincide. Since $m\equiv M_B^TM_B$ is unitary, its eigenvalues have 
absolute value 1 and are bound by $\det m=1$. For such matrices the spectrum is
completely described by a complex number $\tr m$ and a real number
$\tr^2 m-\tr m^2$.

The matrix $m$ is unitary and symmetric. The following statement is used in the 
proof of Theorem 2:

{\sl Lemma.} Any unitary symmetric matrix $m$ has a real orthogonal eigenbasis.

{\sl Proof.} Any eigenbasis of $m$ can be converted into a real orthogonal 
one, as seen from the following observation:
If $\v$ is an eigenvector of $m$ with eigenvalue $\lambda$ then $\v^*$ is also 
an eigenvector with the same eigenvalue (conjugation of 
$m^*\v=m^{-1}\v=\lambda^{-1}\v=\lambda^*\v$ gives this result). Hence, ${\rm 
Re}\v$ and ${\rm Im}\v$ are also eigenvectors.
$\Box$

{\sl Proof of Theorem 2.} 
In the Bell basis local operations transform a unitary gate $M_B$ into 
$O_1M_BO_2$, where $O_1,O_2\in SO(4,{\bf
R})$ are orthogonal matrices.  Therefore $m= M_B^TM_B$ is transformed to 
$O_2^TmO_2$.  Obviously, the spectrum of $m$ is invariant under this 
transformation.

To prove completeness of the set of invariants, we notice that the
lemma above implies that $m$ can be diagonalized by an orthogonal
rotation $O_M$, i.e.  $m=O_M^T d_M O_M$ where $d_M$ is a diagonal matrix.
Suppose that another gate $L$ is given, and $l\equiv L_B^TL_B$ has the same
spectrum as $m$.  Then the entries of $d_M$ and $d_L$ are related
by a permutation.  Hence  $d_M=P^Td_LP$, where $P$ is an orthogonal matrix,
which permutes the basis vectors.  Using the relation of $m$ to
$d_M$ and of $l$ to $d_L$, we conclude that $l=O^TmO$ where $O\in
SO(4,{\bf R})$.

Single-bit operations $O$ and $O'\equiv L_BO^TM_B^{-1}$ transform one gate into 
the other: $L_B=O'M_BO$.
The gate $O'$ is a single-bit operation since it is real and orthogonal.
Indeed, $O'^TO'=(M_B^{-1})^T O L_B^T L_B O^T M_B^{-1}=
(M_B^{-1})^T O l O^T M_B^{-1}= (M_B^{-1})^T m M_B^{-1}= \hat 1$.  On the other 
hand, $O'$ is unitary as a product of unitary
matrices.  This implicates its reality.
$\Box$

So far we have discussed equivalence up to single-bit transformations {\it with
unit determinant}.  However, physically unitary gates are defined only up to an
overall phase factor.  The condition $\det M=1$ fixes this phase factor but not
completely:  multiplication by $\pm i$ preserves the determinant.  With this in
mind, we describe a procedure to verify if two two-qubit unitary gates with
arbitrary determinants are equivalent up to local transformations and an overall
phase factor:

We calculate $m=M_B^TM_B$ for each of them and compare the pairs
$[\tr^2 m\,\det M^\dagger\,;\,\tr m^2\,\det M^\dagger]$.  If they coincide, the 
gates are equivalent and the proof of Theorem 2 allows to express explicitly one 
gate via the other and single-bit gates, $O$ and $O'$.

\paragraph*{Classification of two-qubit states.}

In this section we discuss equivalence and invariants of two-qubit states up to
local operations.  Let us express a 2-bit density matrix in terms of Pauli
matrices acting on the first and the second qubit:  $\hat\rho=\frac{1}{4}\hat 1+
\frac{1}{2}\s\vsigma^1 +\frac{1}{2}\p\vsigma^2 +\beta_{ij}\sigma_i^1\sigma_j^2$.
If the qubits are considered as spin-1/2 particles, then $\s$ and $\p$ are their
average spins in the state $\hat\rho$, while $\hat\beta$ is the spin-spin
correlator:  $\beta_{ij}= \langle S^1_i S^2_j\rangle$.  Any single-bit operation
is represented by two corresponding $3\times3$ orthogonal real matrices of `spin
rotations', $O,P\in SO(3,{\bf R})$.  Such an operation, $O\otimes P$, transforms
$\hat\rho$ according to the rules:  $\s\to O\s$, $\p\to P\p$, and $\hat\beta\to
O\hat\beta P^T$.  We find a set of invariants which completely characterize
$\hat\rho$ up to local gates.

The density matrix is specified by 15 real parameters, while local gates
form a 6-dimensional group.  Thus we expect $15-6=9$ functionally
independent invariants ($I_1$--$I_9$ in Table~\ref{StateInvs}).  However, these
invariants fix a state only up to a finite symmetry group, and
additional invariants are needed~\cite{Grassl}.  In Table I we
present a set of 18 polynomial invariants and prove that the set is complete.
\begin{center}
\parbox{\columnwidth}{%
\begin{table}
\begin{tabular}{c@{\hspace{2mm}}ccc}
$I_{1,2,3}$ & $\det\hat\beta,\; \tr(\hat\beta^T\hat\beta),\; 
		\tr(\hat\beta^T\hat\beta)^2$
& 0 & 0
\\
$I_{4,5,6}$ & $\s^2,\; [\s\hat\beta]^2,\; 
		[\s\hat\beta\hat\beta^T]^2$
& $\s$ & 0
\\
$I_{7,8,9}$ & $\p^2,\; [\hat\beta\p]^2,\; 		
		[\hat\beta^T\hat\beta\p]^2$
& 0 & $\p$
\\
$I_{10}$ & $(\s,\;\s\hat\beta\hat\beta^T,\;\s[\hat\beta\hat\beta^T]^2)$
& $(1;1;\pm1)$ & 0 
\\
$I_{11}$ & $(\p,\;\hat\beta^T\hat\beta\p,\;[\hat\beta^T\hat\beta]^2\p)$
& 0 & $(1;1;\pm1)$ 
\\
$I_{12}$ & $\s \hat\beta\p$
& $(1;\pm b_1^3;0)$ & $(\mp b_2^3;1;0)$
\\
$I_{13}$& $\s \hat\beta\hat\beta^T\hat\beta \p$
& $(1;\pm b_1;0)$ & $(\mp b_2;1;0)$
\\
$I_{14}$ & $e_{ijk}e_{lmn}s_{i}p_{l}\beta_{jm}\beta_{kn}$
& $(0;0;1)$ & $(0;0;\pm1)$
\\
$I_{15}$ & $(\s,\;\s\hat\beta\hat\beta^T,\;\hat\beta\p)$
& $(0;1;1)$ & $(\pm 1;0;0)$
\\
$I_{16}$ & $(\s\hat\beta,\;\p,\;\hat\beta^T\hat\beta\p)$
& $(\pm 1;0;0)$ & $(0;1;1)$
\\
$I_{17}$ & $(\s\hat\beta,\;\s\hat\beta\hat\beta^T\hat\beta,\;\p)$
& $(1;1;0)$ & $(0;0;\pm 1)$
\\
$I_{18}$ & $(\s,\;\hat\beta\p,\;\hat\beta\hat\beta^T\hat\beta\p)$
& $(0;0;\pm 1)$ & $(1;1;0)$
\end{tabular}

\caption{\label{StateInvs}The complete set of invariants of a two-qubit state.
Here $({\bf a},{\bf b},{\bf c})$ stands for the triple scalar product ${\bf
a}\cdot({\bf b}\times{\bf c})$, and $e_{ijk}$ is the
Levi-Cevita symbol; $\s\hat\beta$ is a 3-vector with
components $s_j\beta_{ji}$ etc.
The last two columns present, for each invariant, $\s,\p$ in a
situation when only this invariant distinguishes between two nonequivalent 
states. In these examples we assume a nondegenerate $\hat\beta$, with $b_3=0$ 
for $I_{12-18}$.
}
\end{table}
}
\end{center}

{\sl Theorem 3.} Two states are locally equivalent exactly when the invariants 
$I_1$--$I_{18}$ have equal values for these states. (Only signs of $I_{10}$, 
$I_{11}$, $I_{15-18}$ are needed.)

None of the invariants can be removed from the set without
affecting completeness, as demonstrated by examples in the table.

The proof below gives an explicit procedure to find single-bit gates which 
transform one of two equivalent states into the other.

{\sl Proof.} It is clear that all $I_i$ in the table are invariant under 
independent orthogonal rotations $O$, $P$, i.e. under single-bit gates. To 
prove 
that they form a complete set, we show that for given values of the 
invariants one can by local operations transform any density matrix with these 
invariants to a specific form. In the course of the proof we fix more and more 
details of the density matrix by applying local gates (this preserves the 
invariants).

The first step is to diagonalize the matrix $\hat\beta$, which can be achieved
by proper rotations $O$, $P$ (singular value decomposition).  The invariants
$I_1$--$I_3$ determine the diagonal entries of $\hat\beta$ up to a simultaneous
sign change for any two of them.  Using single-bit operations $\R{i}\otimes\hat
1$ (where $\R{i}$ is the $\pi$-rotation about the axis $i=1,2,3$) we can fix
these signs:  all three eigenvalues, $b_1, b_2, b_3$, can be made nonnegative,
if $I_1=\det\hat\beta\ge0$, or negative, if $\det\hat\beta<0$.  Further
transformations, with $O=P$ representing permutations of basis vectors, place
them in any needed order.  From now on we consider only states with a fixed
diagonal $\hat\beta$.  Hence, only such single-bit gates, $O\otimes P$, are
allowed which preserve $\hat\beta$.  The group of such operations depends on
$b_1,b_2,b_3$.  We examine all possibilities below, showing that the invariants
$I_{4-18}$ fix the state completely.

(A) $\hat\beta$ is nondegenerate: all $b_i$ are different.
Then the only $\hat\beta$-preserving local gates are 
$\r{i}\equiv\R{i}\otimes\R{i}$. 

The invariants $I_{4-6}$ and $I_{7-9}$ set absolute values of six components 
$s_i$, $p_i$, but not their signs. The signs are bound by other invariants. In 
particular, $I_{10}$ and $I_{11}$ fix the values of $s_1s_2s_3$ and 
$p_1p_2p_3$.
Furthermore, $I_{12-14}$ give three linear constraints on three quantities 
$s_ip_i$. When $\hat\beta$ is not degenerate, one can solve them for $s_ip_i$.
Let us consider several cases:

i) $\s$ has at least two nonzero components, say $s_1,s_2$. These two can be 
made positive by single-bit gates $\r{1}$, $\r{2}$. After that, the signs of 
$p_{i=1,2}$ are fixed by values of $s_ip_i$, while $I_{10}$ fixes the 
sign of $s_3$.
The sign of $p_3$ can be determined from $s_3p_3$, if $s_3\ne0$;
if $s_3=0$ then $p_3$ is fixed by $I_{15}=p_3s_1s_2 b_3(b_2^2-b_1^2)$ or
$I_{17}=p_3s_1s_2b_1b_2(b_2^2-b_1^2)$.

If $\p$ has at least two nonzero components, a similar argument applies, with
$I_{11,16,18}$ instead of $I_{10,15,17}$.  If both $\s$ and $\p$ have at most
one nonzero component, $s_i$ and $p_j$, then either ii) $i=j$, and the signs
are specified by $s_ip_i$ up to $\hat\beta$-preserving gates $\r{k}$; or iii)
$i\ne j$, and one can use $\r{i},\r{j}$ to make both components nonnegative.

(B) $\hat\beta$ has two equal nonzero eigenvalues: $b_3\ne b_1=b_2\ne0$.
We define horizontal components $\s_\perp=(s_1,s_2,0)$, 
$\p_\perp=(p_1,p_2,0)$.
Then $\hat\beta$-preserving operations are generated by simultaneous, 
coinciding 
rotations of $\s_\perp$ and $\p_\perp$ [i.e. $\s_\perp\to O\s_\perp$, 
$\p_\perp\to O\p_\perp$, where $O\in SO_{1,2}(2)$ is a 2D-rotation], as well as 
$\r{i}$.

The invariants $I_{4-9}$ fix $\s_\perp^2$, $s_3^2$, $\p_\perp^2$ and $p_3^2$. 
To 
specify $\s$ and $\p$ completely the angle between $\s_\perp$ and $\p_\perp$, 
as 
well as the signs of $s_3$ and $p_3$ should be determined. 
These are bound by the remaining invariants. In particular, $I_{12-14}$ fix 
$s_3p_3$ and $\s_\perp\p_\perp$.
The latter sets the angle between $\s_\perp$ and $\p_\perp$ up to a sign.

There are two possibilities:
i) $s_3=p_3=0$. The states with opposite angles are related by $\r{1}$ and 
hence 
equivalent.
ii) $s_3\ne0$ (the case $p_3\ne0$ is analogous).  Applying $\r{1}$, if needed, 
we can assume that $s_3$ is
positive.  Then, $p_3$ is specified by the value of $s_3p_3$.  Apart from that,
$I_{15}$ sets $(\s_\perp\times\p_\perp)_3s_3$, and hence the sign of the cross 
product $\s_\perp\times\p_\perp$.  This fixes the density matrix completely.

(C) $b_1=b_2=0$, $b_3\ne0$.  In this case $\hat\beta$-preserving operations are
independent rotations of $\s_\perp$ and $\p_\perp$ [which form
$SO_{1,2}(2)^{\otimes2}$] and $\r{i}$.  The invariants $I_{4-9}$ fix
$\s_\perp^2$, $s_3^2$, $\p_\perp^2$, and $p_3^2$, while $I_{12}$ (or $I_{13}$)
sets $s_3p_3$.  It is easy to see that they specify the state completely.

(D) $b_1=b_2=b_3\ne0$.
All transformations $O\otimes O$, where $O\in SO(3)$, preserve $\hat\beta$.
The invariants $I_4$, $I_7$ and $I_{12}$ fix $\s^2$, $\p^2$ and $\s\p$, and 
this 
information is sufficient to determine $\s$ and $\p$ up to a rotation.

(E) $\hat\beta=0$.
In this case all local transformations preserve $\hat\beta$;
hence, $\s$ and $\p$ can rotate independently. The only nonzero invariants, 
$I_4$ and $I_7$, fix $\s^2$ and $\p^2$.
$\Box$

\paragraph*{Discussion.}

In this section we demonstrate applications of our results.
We calculate the invariants for several two-qubit gates to find out which of 
them are locally equivalent. These gates include CNOT, SWAP and its square 
root, 
as well as several gates $\exp(i{\cal H}t)$ generated by hamiltonians 
$\frac{1}{4}\vsigma^1\vsigma^2$, $\frac{1}{4}\vsigma^1_\perp\vsigma^2_\perp$
[here $\vsigma_\perp=(\sigma_x;\sigma_y)$] and
$\frac{1}{4}\sigma^1_y\sigma^2_y$ (cf.~Refs.~\cite{Our_Nature,Loss,LossOptim}) 
after evolution during time $t$.
\begin{center}
\parbox{\columnwidth}{%
\begin{table}
\begin{tabular}{ccccc}
&Identity&CNOT&SWAP&$\sqrt{\rm SWAP}$\\
$G_1$&1&0&-1&$i/4$\\
$G_2$&3&1&-3&0\\
\hline\hline
\\[-3mm]
$\cal H$& $\vsigma^1\vsigma^2$ &
$\vsigma^1_\perp\vsigma^2_\perp$ &
$\sigma^1_y\sigma^2_y$ &
\\[1mm]
\hline
\\[-1mm]
$G_1$ &
$\frac{1}{16}e^{it}(3+e^{-2it})^2$ &
$\cos^4(t/2)$ &
$\cos^2(t/2)$
\\[1mm]
$G_2$& $3\cos t$ &
$1+2\cos t$ & $2+\cos t$
\end{tabular}
\caption{\label{GateInvs}Invariants of two-qubit gates, $G_1=\tr^2 m/16\det M$ 
and $G_2=(\tr^2m-\tr m^2)/4\det M$. The latter is always a real number.
Numerical prefactors are chosen to simplify expressions.
}
\end{table}
}
\end{center}
Analyzing the invariants we see that to achieve CNOT one needs to perform a 
two-qubit gate at least twice if the latter is triggered by the
Heisenberg hamiltonian $\vsigma\vsigma$. At the same time SWAP and $\sqrt{\rm 
SWAP}$ can be performed with one elementary two-bit gate~\cite{LossOptim}. The 
interaction $\sigma_y\sigma_y$ allows to perform CNOT in one step, while 
$\vsigma_\perp\vsigma_\perp$ requires at least two steps for all three gates.

In Josephson qubits~\cite{Our_Nature} elementary two-bit gates are generated by
the interaction ${\cal H}=-\frac{1}{2}E_{\rm J}
(\hat\sigma_x^1+\hat\sigma_x^2)+
(E_{\rm J}^2/E_L) \hat\sigma_y^1\hat\sigma_y^2$.
Investigation of the invariants shows that CNOT can be performed if $E_{\rm J}$ 
is tuned to $\alpha E_L$ for a finite time $t=\alpha\pi(2n+1)/4E_L$, where $n$ 
is an integer and $\alpha$ satisfies 
$\alpha^2\cos[\pi(n+\frac{1}{2})\sqrt{1+\alpha^{-2}}]=-1$.

For creation of entanglement between qubits a useful property of a two-qubit
gate is its ability to produce a maximally entangled state (with
$|\Ent\psi|=1/2$) from an unentangled one~\cite{LossOptim}.  This property is
locally invariant and one can show that a gate $M$ is a perfect entangler
exactly when the convex hull of the eigenvalues of the corresponding matrix $m$
contains zero.  In terms of the invariants, introduced in Table~\ref{GateInvs},
this condition reads:  $\sin^2\gamma\le4|G_1|\le1$ and
$\cos\gamma(\cos\gamma-G_2)\ge0$, where $G_1=|G_1|e^{i\gamma}$.  Among the
gates in the table CNOT and $\sqrt{\rm SWAP}$ are perfect entanglers.  The
Heisenberg hamiltonian can produce only two perfect entanglers ($\sqrt{\rm
SWAP}$ or its inverse), while $\sigma_y\sigma_y$ - only CNOT.  At the same
time, the interaction $\vsigma_\perp\vsigma_\perp$ produces a set of perfect
entanglers if the system evolves for time $t$ with $\cos t\le0$.

To conclude, we have presented complete sets of local polynomial invariants of
two-qubit gates (3 real invariants) and two-qubit mixed states (18 invariants) 
and demonstrated how these results can be used to optimize quantum logic
circuits and to study entangling properties of unitary operations.

I am grateful to G.~Burkard, D.P.~DiVincenzo, M.~Grassl, G.~Sch\"on, and 
A.~Shnirman for fruitful discussions.

\end{document}